\documentclass[twocolumn,showpacs,amsmath,amssymb]{revtex4}
\usepackage{graphicx}
\usepackage{color}
\usepackage{amsmath}
\usepackage{amssymb}
\usepackage{booktabs}
\usepackage{float}

\begin{document}
\title{Relating Microstructure and Particle-level Stress in Colloidal Crystals Under Increased Confinement}
\author{Neil Y.C. Lin}
\affiliation{Department of Physics, Cornell University, Ithaca, New York 14853}
\author{Itai Cohen}
\affiliation{Department of Physics, Cornell University, Ithaca, New York 14853}

\begin{abstract}
The mechanical properties of crystalline materials can be substantially modified under confinement. Such modified macroscopic properties are usually governed by the altered microstructures and internal stress fields. Here, we use a parallel plate geometry to apply a quasi-static squeeze flow crushing a colloidal polycrystal while simultaneously imaging it with confocal microscopy. The confocal images are used to quantify the local structure order and, in conjunction with Stress Assessment from Local Structural Anisotropy (SALSA), determine the stress at the single-particle scale. We find that during compression, the crystalline regions break into small domains with different geometric packing. These domains are characterized by a pressure and deviatoric stress that are highly localized with correlation lengths that are half those found in bulk. Furthermore, the mean deviatoric stress almost doubles, suggesting a higher brittleness in the highly-confined samples.
\end{abstract}

\maketitle

\section{Introduction}
Understanding the effect of confinement on crystalline materials is crucial to many technological applications, such as lubrication, adhesion, and fabricating novel optical, catalytic materials~\cite{alba2006effects, dimitrov1996continuous, jiang1999single, wasan2003spreading, abkarian2004colloidal}. For example, during film processes,  which are important for optoelectronics, magnetic and electronic materials~\cite{diyatmika2015thin, tsai1987characterization, park200842, akimoto2006thin}, internal stresses often arise affecting the mechanical stability and strength of the film~\cite{abermann1985internal, thornton1989stress, freund2004thin, akimoto2006thin, harris2007patterning, harris2009evaporative, singh2007cracking, shereda2008local, xu2013imaging, koch1994intrinsic}. Here we study increasingly confined colloidal crystals to investigate how changes in their microstructure affect the microscopic stresses ultimately responsible for their bulk mechanical properties. Such colloidal systems are comprised of particles small enough to undergo Brownian motions that preserve a thermodynamic ensemble, while still large enough to be optically imaged with high temporal and spatial resolutions~\cite{derks2009dynamics, van1992three, prasad2007confocal, van1997template, verhaegh1995direct, royer2015rheology, conrad2006contribution}. As such, colloidal crystals have been widely used as a model system to elucidate many phenomena associated with confinement including crystal growth~\cite{kumacheva2002colloid}, packing~\cite{fortini2006phase, nagamanasa2011confined, pieranski1983thin, cohen2004shear, neser1997finite}, and melting mechanisms\cite{peng2010melting}. 

In principle, knowing all particle positions and their interactions is sufficient to determine the suspension's structural order and stress distribution. Unfortunately, resolving the stress distribution within colloidal materials has remained experimentally challenging. In the simplest colloidal system -- hard-spheres -- particles do not interact until contact. The steep hard-sphere potential and experimental noise in locating particles make potential-based stress calculation (force times relative vector) impractical in experiment. In this work, we use a new technique -- Stress Assessment from Local Structural Anisotropy (SALSA) -- to measure the stress distribution in confined hard-sphere colloidal polycrystals~\cite{lin2016measuring}. SALSA uses the particle positions captured using confocal microscopy to calculate the orientation dependent particle collision probabilities. These probabilities are used to determine the stress at the single particle scale. Using SALSA we follow the evolving stress distribution in a polycrystal as it is compressed quasi-statically between two parallel plates.

\section{Experiments}\label{exp}
\subsection{Samples and instruments}
We conduct experiments with suspensions comprised of sterically stabilized Poly(methyl methacrylate), PMMA, particles. The particles have a diameter of $2a=1.62$ $\mu$m, polydispersity $\approx3\%$, and are fluorescently labeled with DiIC$_{18}$(3) (1,1'-Dioctadecyl-3,3,3',3'-Tetramethylindocarbocyanine Perchlorate) for confocal imaging. The solvent -- a mixture of decalin and CXB (cyclohexylbromide) -- has a refractive index and density that nearly match those of the PMMA particles. While a slight mismatch between the particle and solvent density $\Delta \rho \approx$ 0.03 g/cm$^3$ is introduced to induce slow sedimentation for the bulk samples, the confined crystal samples are all density-matched. We approximate the hard-sphere interparticle force by saturating the solvent with tetrabutyl ammonium bromide (TBAB) at a concentration ~($\approx$ 260 nM)~\cite{royall2006re}, and using a syringe filter to remove any excess salt granules~\cite{yethiraj2003colloidal}. The added TBAB salt screens the electrostatic force and results in a Debye length, $\approx$ 100 nm, substantially smaller than the particle diameter\cite{royall2013search}.

We use a high-speed confocal laser scanning microscope (Zeiss, LSM 5 LIVE) to image the 3D structure of the sample. We acquire a time series of 15 image stacks where each stack contains up to 512$\times$512$\times$500 voxels, corresponding to a sample volume of 71$\times$71$\times$68 $\mu$m$^3$. We use our confocal rheoscope to confine our crystal samples between two plates with a separation that is uniform to within $\pm 0.2$ $\mu$m~\cite{lin2013far, lin2014biaxial, lin2014multi}. The top plate, a silicon wafer (4 mm $\times$ 4 mm), is fixed to the rheoscope's kinematic mount and is static throughout the experiment. The bottom plate, a transparent coverslip for imaging from below, is attached to a multi-axial piezo electric, and can be moved vertically ($z$-axis) to change the gap size~\cite{lin2014multi}. The multi-axial piezo (PI P-563.3CD) has a travel range 300 $\mu$m in the $z$ direction and an accuracy $\pm$ 2 nm. Using this parallel-plate setup we reduce the gap heights in a controlled manner from 38 $\mu$m down to 6 $\mu$m (24$\ge h/2a \ge 4$).

\subsection{Structure and Stress Measurements}\label{measure}
We process the confocal data and locate the particle positions using the Crocker-Grier featuring algorithm~\cite{crocker1996methods} that locates particle positions with sub-pixel accuracy ($\approx 50$ nm)~\cite{dinsmore2001three, prasad2007confocal, besseling2007three, ramsteiner2009experimental}. From the particle positions, we employ a standard bond-order parameter method~\cite{hernandez2009equilibrium, ten1996numerical, steinhardt1983bond, gasser2001real, dullens2006dynamic} to calculate the local three dimensional (3D) structure order. In the bond-order parameter calculation, we determine the normalized complex order parameter $\hat q_{lm}(\alpha)$ for each particle $\alpha$, $\hat q_{lm}(\alpha) = \frac{1}{C}\langle Y_{lm}(\hat r_{\alpha \beta})  \rangle_{\beta \in nn}$, where factor $C$ normalizes the order parameter such that $\sum_m \hat q_{lm}(\alpha)  \hat q^*_{lm}(\alpha) =1$, $Y_{lm}(\hat r_{\alpha \beta})$ is the spherical harmonic function of the unit vector $\hat r_{\alpha \beta}$ pointing from particle $\alpha$ to $\beta$, and $\langle \ldots \rangle_{\beta \in nn}$ denotes the average of the neighbors of particle $\alpha$. The neighboring particles are defined as those with a center-center distance within $1.41(2a)$, which coincides with the first minimum of the radial distribution function $g(r)$. Following previous protocol, we set $l=6$~\cite{hernandez2009equilibrium, ten1996numerical, steinhardt1983bond, gasser2001real, dullens2006dynamic}. The number of ordered neighbors is then approximated by summing the complex inner product, $N_{\rm{ord}} = \sum_{m,\beta} \hat q_{lm}(\alpha)  \hat q^*_{lm}(\beta)$. Here, the number of ordered neighbors has a range $0 \le N_{\rm{ord}} \le12$. \footnote{This range is identical to those calculated in the previously employed threshold method, $N_{\rm{ord}}^\prime = \sum_m \mathcal{H}(\hat q_{lm}(\alpha)  \hat q^*_{lm}(\beta)-0.5)$ where $\mathcal{H}$ denotes Heaviside function~\cite{hernandez2009equilibrium}. Compared to the previous definition, $N_\mathrm{ord}$ reports fractional contribution to the local ordering, and more finely resolves the crossover between crystalline and defect structures.}

From the particle positions, we can also measure the stress at the single particle scale using SALSA. In our Brownian hard-sphere systems, the force with which particles collide is related to the thermal energy $k_B T$. Using a time series of featured particle positions, we deduce the thermal collision probability, and compute the stress arising from these collisions. As shown by previous work~\cite{lin2016measuring}, the stress tensor $\sigma^\alpha_{ij} = \sigma_{ij}(\vec X^\alpha)$ at particle $\alpha$ can be approximated by $\sigma^{\alpha}_{ij} =  \frac{k_B T}{\Omega^\alpha}\left(\frac{a}{\Delta}\right)  \langle \psi^{\alpha}_{ij} (\Delta) \rangle$, where $\Omega^{\alpha}$ is the volume occupied by the particle, $\Delta$ is the cutoff distance from contact. Here, $\langle \psi^{\alpha}_{ij}(\Delta) \rangle$ is the time-averaged {\em local structural anisotropy} for the particle $\alpha$, $\langle \psi^{\alpha}_{ij}(\Delta) \rangle=\langle \sum_{\beta \in nn} \hat{r}^{\alpha\beta}_{i}\hat{r}^{\alpha\beta}_{j} \rangle$, where $nn$ denotes the particles that lie within a distance $2a+\Delta$ from particle $\alpha$, $ij$ are spatial indices, and $\hat{r}^{\alpha\beta}$ is the unit vector pointing from particle $\alpha$ to particle $\beta$. 

In granular literature, $\langle \psi^{\alpha}_{ij}(\Delta) \rangle$ also denotes time-averaged fabric tensor~\cite{bi2011jamming}. While the trace of the time-averaged fabric tensor $\sum_i \psi^\alpha_{ii}$ determines the contact particle number, the off-diagonal terms report the anisotropy of these contact particle configurations. When averaged over time, the fabric tensor of the selected particle captures the probability of Brownian collisions between it and its neighbors. This probability is linearly proportional to the cutoff distance, or shell thickness $\Delta$ when $\Delta \ll a$. Therefore, when scaling the collision probability by $\Delta$, the stress is independent of $\Delta$.

The last step of the SALSA calculation scales the probability by the energy density per collision $k_B T/\Omega^\alpha$. In a defect-free crystal, $\Omega^\alpha$ is simply the system volume divided by the particle number. However, in a crystal containing defects, the local volume occupied by each particle varies. This variation needs to be considered to correctly measure the stress near defects. So, we first calculate the pointwise stress, $\sigma_{ij}^{pt}(\vec x)  = \frac{k_B T}{dV} \frac{a}{\Delta} \langle \psi_{ij}(\vec x, \Delta) \rangle$ then perform a spatial average to obtain a macroscopic measurement at the particle-level~\cite{tadmor2011modeling}. Here, $dV$ in theory, should be an infinitesimal volume. In practice, when the pointwise stresses are assigned to a 3D discrete grid, $dV$ is the single box volume of the grid. The continuum stress field $\sigma_{ij}^{cont}(\vec x)$ is then smoothed, ${\sigma}^{cont}_{ij}(\vec{x}) = \int_{\vec{y} \in R} w(\vec y - \vec x) {\sigma}_{ij}^{pt}(\vec{x}) d\vec y$, where in our experiments $w(r)$ -- the weighting kernel -- is a Gaussian function $\pi ^{-3/2} r_w^{-3} e^{-\frac{r^2}{2r_w^2}}$~\cite{tadmor2011modeling}. In particular, we set $r_w = 2a$ to remove stress features on length scales smaller than a particle. We have tested different discrete grid sizes, and other normalized weighting kernels, and find the results insensitive to those changes. This spatial average effectively addresses local volume variation. ~Finally, while SALSA can be modified to account for the contributions due to the confining surfaces, for simplicity, in the data presented here, we exclude boundary particles in all final presentations of local structure and stress distributions.

\section{Bulk Crystals}\label{bulk}

\begin{figure*}[htp]
\includegraphics[width=0.7 \textwidth]{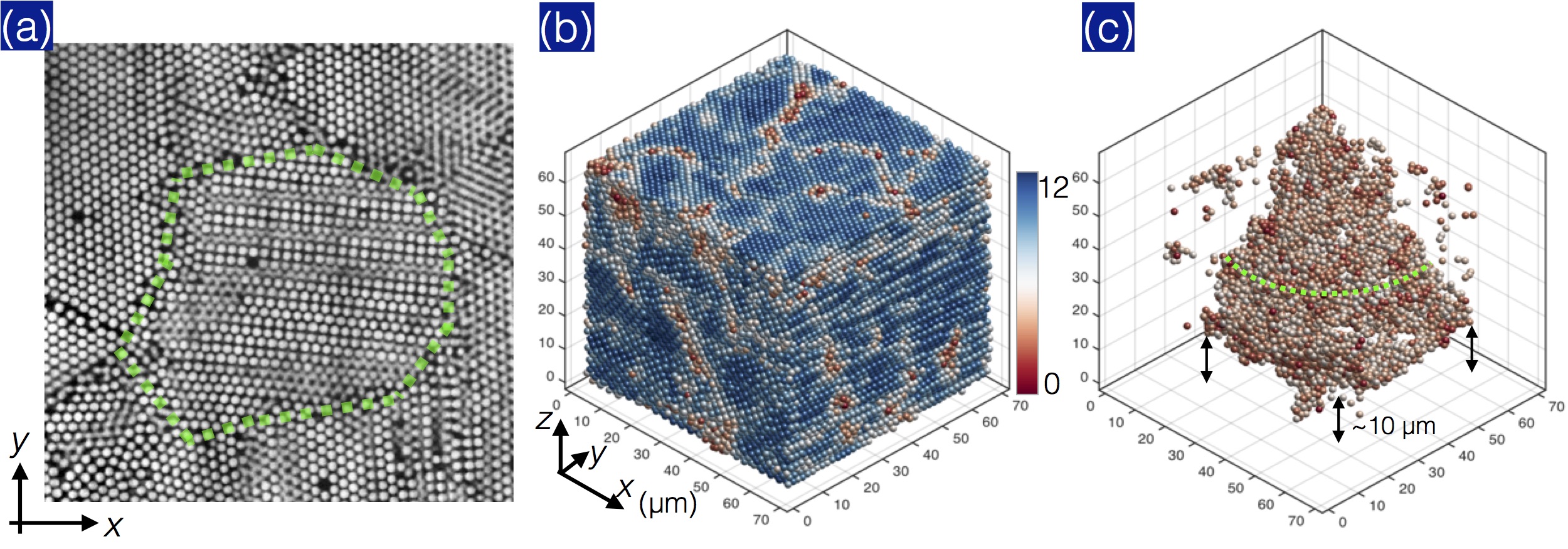}
\caption{Microstructure and local structure order of the bulk PMMA polycrystal. (a) The $\hat x \hat y$ slice of confocal data shows the particle configuration. The grain boundaries surrounding the central crystallite are highlighted by green dashed lines. The 3D distribution of order neighbor number $N_{\mathrm{ord}}$ is shown in (b). In contrast to the crystalline domains (blue particles), the defect regions (red particles) have lower values of $N_{\mathrm{ord}}$. By thresholding $N_{\mathrm{ord}}$ (c), we uncover the particles following the grain boundaries highlighted in (a) with green dashed lines.}
\label{nfig_1_structure}
\end{figure*}

In the tested bulk crystal, we observe different types of defects including vacancies, dislocations, stacking faults, grain boundaries, and voids. These defects are often close and can interact with one another. The typical size of a single crystallite is approximately 50 $\mu$m $\times$ 50 $\mu$m $\times$ 50 $\mu$m, containing on the order of $10^3$ particles. We show a horizontal slice of a 3D confocal image in Fig.~\ref{nfig_1_structure}(a). In the image, we see a crystallite in the center surrounded by several other domains. The grain boundary of the center domain is highlighted by the dashed red contour. By performing the bond-order parameter analysis, we determine the number of ordered neighbor particles $N_{\rm{ord}}$, and plot its distribution in Fig.~\ref{nfig_1_structure}(b). The red particles represent defect regions with lower $N_{\rm{ord}}$; blue particles are crystalline domains with higher $N_{\rm{ord}}$. To better visualize defects deeply embedded in the crystal, we remove crystalline particles with $N_{\rm{ord}} > 5.5$. The remaining particles are shown in Fig.~\ref{nfig_1_structure}(c). For further clarity, we remove particles with $z$-positions higher than 60 $\mu$m. We find that the thresholded particle distribution faithfully captures the grain boundaries highlighted in the raw confocal image illustrated by Fig.~\ref{nfig_1_structure}(a). 

\begin{figure} [htp]
\includegraphics[width=0.4 \textwidth]{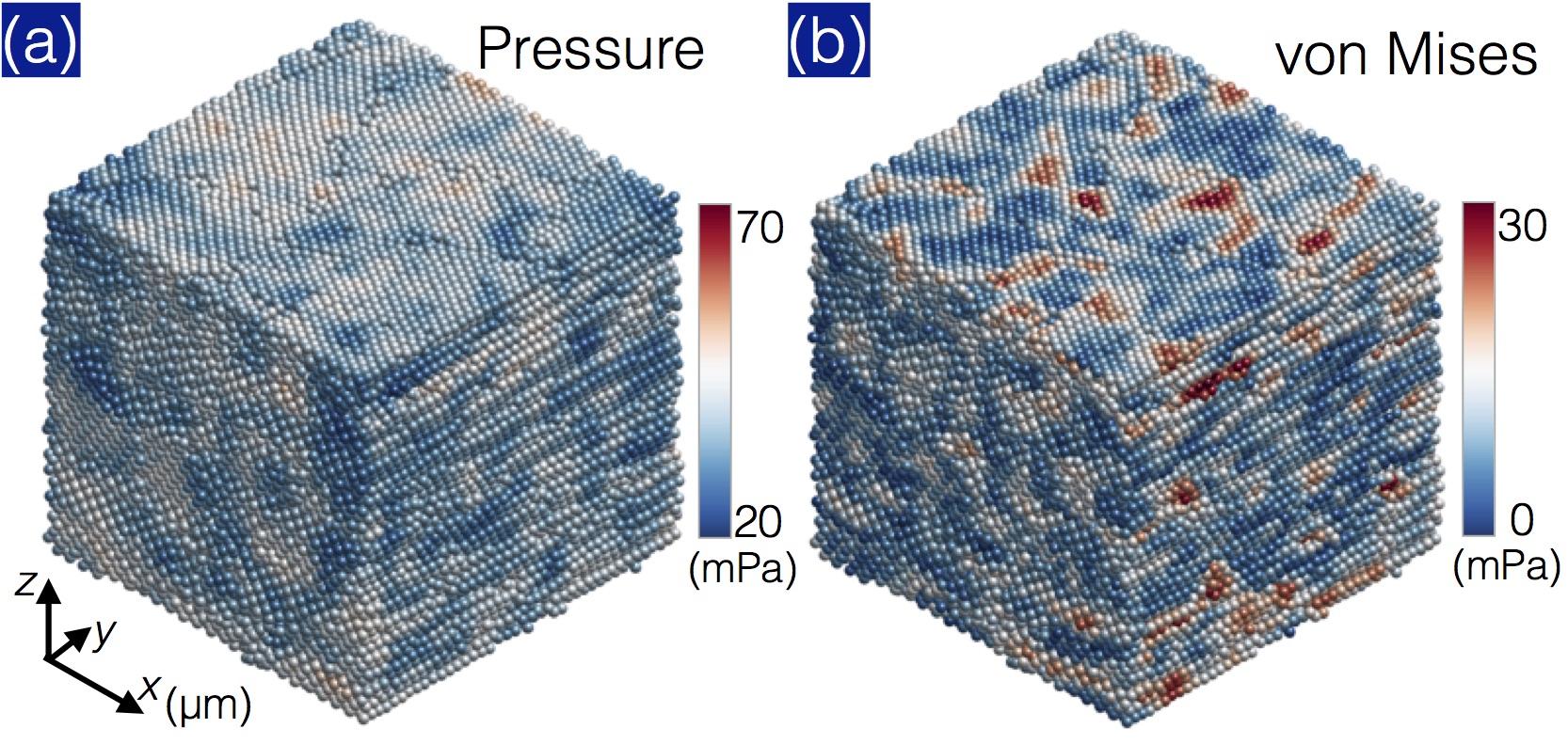}
\caption{3D distributions of pressure and von Mises stress of the bulk polycrystal. (a) Pressure distribution. (b) von Mises stress. The color bars denote the magnitude of the stress. The stress distribution determined by SALSA is a continuous field. We then resample the stress data at individual particle positions, and plot them in the same fashion as local structure data shown in Fig.~\ref{nfig_1_structure}(b) for comparison.}
\label{nfig_2_pressure_vonMises}
\end{figure}

To determine whether the crystalline order affects the stress distribution we apply SALSA to our polycrystal. Since the magnitudes of individual stress components depend on the orientation of the coordinate system, we focus on primary tensor invariants, including the pressure, the three principal stresses, and the von Mises stress. Our polycrystal sample has a mean pressure $\bar P = \frac{1}{3} (\bar \sigma_{xx}+\bar \sigma_{yy} + \bar \sigma_{zz})$ approximately 39 mPa, consistent with the prediction from previous numerical simulations~\cite{pronk2003large}. Furthermore, the measured pressure as a function of height is also consistent with the estimated trend of hydrostatic pressure arising from gravitational settling due to the slight mismatch in the particle and solvent densities ($\Delta \rho \approx$ 0.03 g/cm$^3$). See Sec.~\ref{bulk_pressure} for detailed analyses on the bulk pressure.

In Fig.~\ref{nfig_2_pressure_vonMises}(a) we show the 3D pressure field of our polycrystal sample. The red particles indicate regions with higher pressure and blue particles indicate regions with lower pressure. The pressure fluctuation has a length scale $\sim$ 10 particles considerably smaller than the size of a single domain indicating  \textit{intragrain} stress fluctuations. By comparing the pressure (Fig.~\ref{nfig_2_pressure_vonMises}(a)) and $N_{\rm{ord}}$ (Fig.~\ref{nfig_1_structure}(b)) distributions, we find that the pressure fluctuation has a relatively random spatial distribution, and is not notably correlated with the grain boundary arrangement. 

We also determine the three principal stresses, $\sigma_1$, $\sigma_2$, and $\sigma_3$, by calculating the eigenvalues of the measured three-by-three stress matrix, see Sec.\ref{bulk_principal_vonMises}. We find that the distribution of each principle stress is similar to that of the pressure. We therefore extract the difference between them and calculate the deviatoric -- von Mises stress $\sigma_{\rm{VM}} = \sqrt{\frac{1}{2} \Big[ (\sigma_1-\sigma_2)^2 + (\sigma_2-\sigma_3)^2 + (\sigma_3-\sigma_1)^2 \Big]}$. Similar to pressure, $\sigma_{\rm{VM}}$ is also an invariant. Since this invariant satisfies the property that two stress configurations with equal deviatoric strain energy have the same value of $\sigma_{\rm{VM}}$, such a scalar is effectively a stress field signature capturing the distortion energy of a material under various loads~\footnote{Consider a pure shear case where a metal yields at a critical shear stress $\sigma^c_{xy}$: the corresponding von Mises stress is $\sigma_{\rm{VM}} = \sqrt 3 \sigma^c_{xy}$. Therefore, if a material yields in uniaxial tension at $\sigma_{\rm{VM}} = \sigma^c_{xx}$, it will also yield under a shear strain, but now at a lower value ($\sigma^c_{xy} = 1/\sqrt 3 \sigma^c_{xx}$).}. In continuum elasticity, the mean of $\sigma_{\rm{VM}}$ is often used as a criterion in determining how close a metal or alloy sample is to the yield point when subjected to loads~\cite{schuh2007mechanical}. In atomic-scale simulations, similar yielding criteria have been employed to elucidate the local yielding mechanism in glassy systems\cite{schuh2003atomistic}.

In Fig.~\ref{nfig_2_pressure_vonMises}(b), we show the von Mises stress field $\sigma_{\rm{VM}}$ of our bulk polycrystal. We find that the $\sigma_{\rm{VM}}$ distribution is also roughly random. Counterintuitively, the distribution of the high $\sigma_{\rm{VM}}$ particles does not follow the trend of the grain boundaries shown in Fig.~\ref{nfig_1_structure}(a). In fact, many of the highly stressed regions are well within the crystalline domains (see Sec.\ref{bulk_principal_vonMises} and Fig.~\ref{afig_4_threshold_vonMises}). The observed uniform and random fluctuations of $P$ and $\sigma_{\rm{VM}}$ are consistent with the evenly spread stress fluctuations previously reported in polycrystals of hard-sphere silica particles~\cite{lin2016measuring}. 

\section{Confined Crystals}

\begin{figure*}[htp]
\includegraphics[width=0.75 \textwidth]{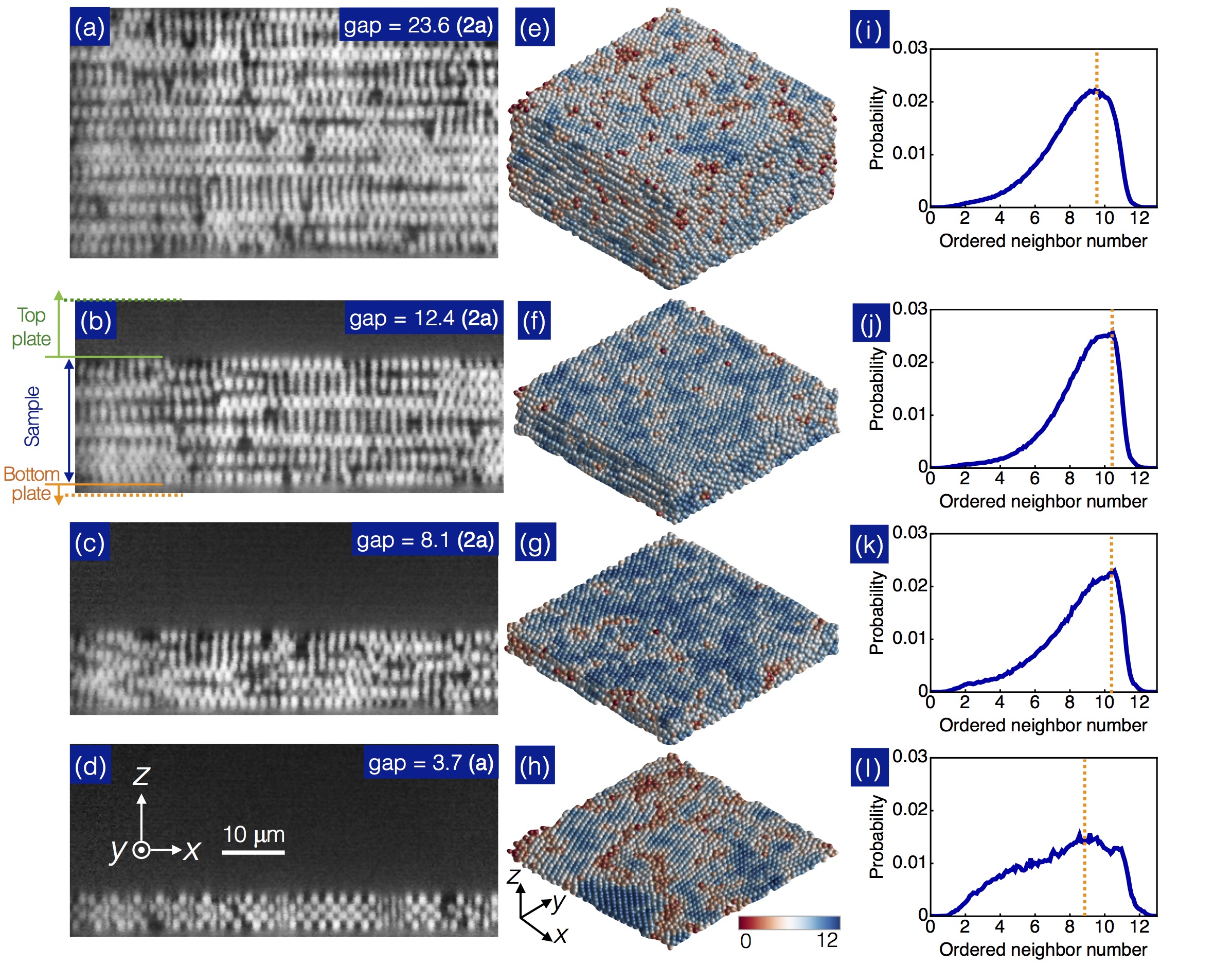}
\caption{Local structure measurements of confined polycrystals. (a-d) Orthogonal slices of confocal images for four representative gap sizes $h$ (labeled).  The sample is confined by a silicon wafer (top) and a coverslip (bottom). The plates are not fluorescently labeled, so they appear dark. The coordinate definition and scale bar are shown in the inset of (d). (e-f) Distributions of ordered neighbor number shown in the same (a-d) order. The top and bottom particle layers are removed in the final presentation. In the ultra-confined sample (h = 3.7(2a)), we observe small crystallines with different packing phases due to geometric constraint. The boundaries between these ordered domains have disordered packing configurations as indicated by the red particles. (i-l) Histograms of ordered neighbor number $N_{\rm{ord}}$ are shown in the same order. The orange dashed line indicates the evolution of the most probable $N_{\rm{ord}}$. Overall, the local structure order increases slightly with decreasing gap height when $h\ge 10(2a)$, then reduces with decreasing gap height when $h< 10(2a)$.}
\label{fig7_confined_bond_order}
\end{figure*}

Confinement is known to affect the structure of crystals~\cite{pieranski1983thin, cohen2004shear, neser1997finite, fortini2006phase, schmidt1996freezing, han2008geometric, de2015entropy, dziomkina2005colloidal}. Here, we use SALSA to determine whether such structure modifications are accompanied by changes in the stress distribution. To create a confined polycrystal, we load our crystalline suspension -- volume fraction $\phi\approx 0.63$ -- in a confocal rheoscope. The rheoscope has three differential screws that allow us to finely adjust the alignment and gap $h$ between two parallel plates holding the sample. We study the confined sample for seven different gap heights, starting with a bulk sample ($h = 38$ $\mu$m $\approx 24 (2a)$). We then gradually move the bottom plate upward, reducing the gap height down to $h = 6$ $\mu$m $\le 4 (2a)$. When the gap decreases, the parallel plates compress the sample and induce a squeeze flow that drives particles outward causing additional structural rearrangement. An oscillatory shear flow with a small strain amplitude $10\%$ and frequency of $1$ Hz along $\hat x \hat z$ is applied for 200 cycles to prevent local jamming and speed up sample recrystallization. Importantly, this low strain amplitude does not generate large structural rearrangements. Thus, the final structure at each height $h$ is primarily determined by the squeeze flow and degree of confinement.

\subsection{Structure}

\begin{figure} [htp]
\includegraphics[width=0.4 \textwidth]{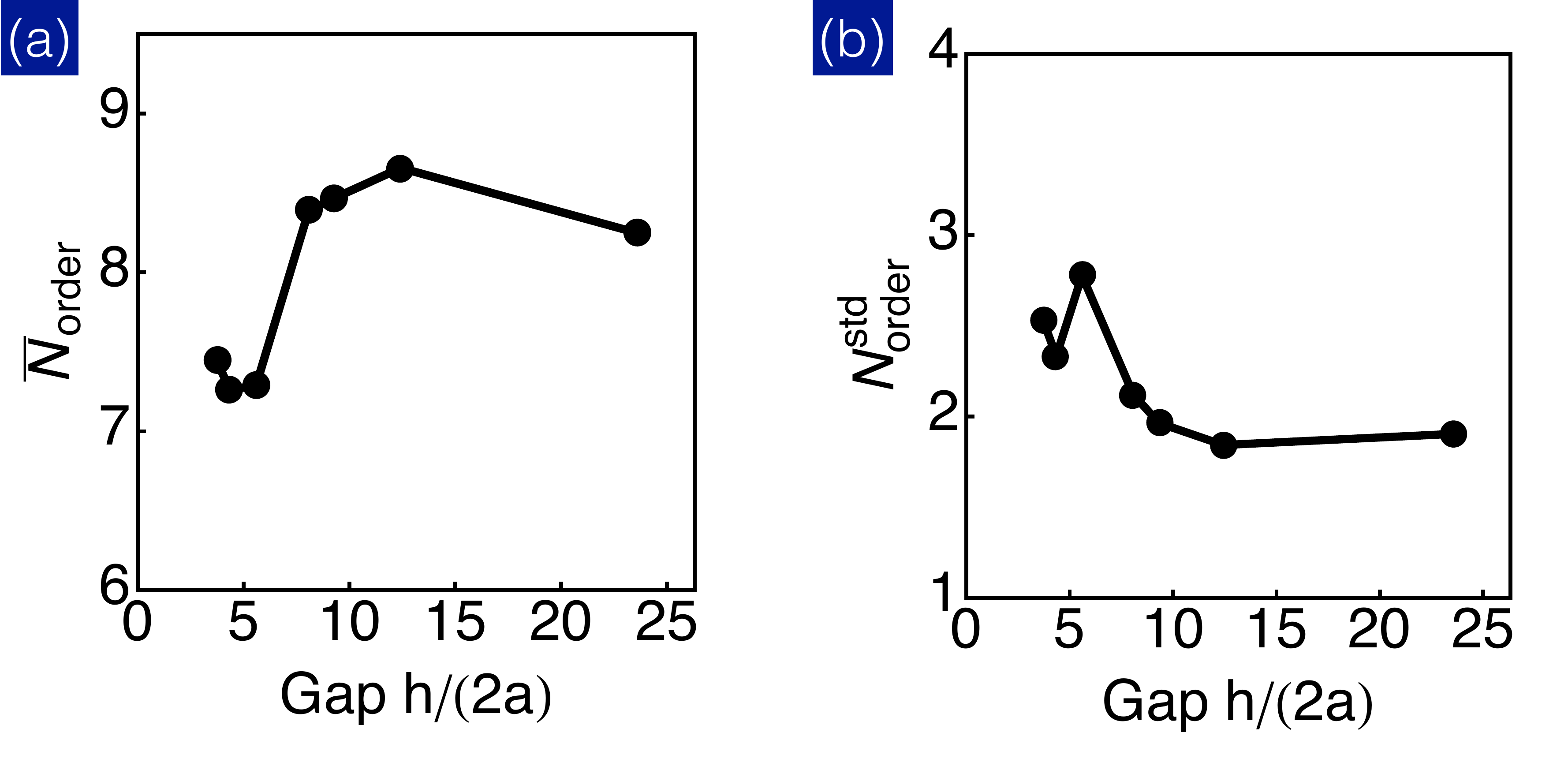}
\caption{The statistics of ordered neighbor number $N_{\rm{ord}}$ in confined samples. (a) Mean value $\bar N_{\rm{ord}}$ slightly peaks around $h/2a \sim 13$ then decays with decreasing $h$. (b) Standard deviation $N^{\rm{std}}_{\rm{ord}}$ remains roughly unchanged when $h/2a \ge10$, and starts to increase when $h/2a < 10$.}
\label{fig8_ordered_number}
\end{figure}

We show representative orthogonal slices from the confocal images for four gaps in Fig.~\ref{fig7_confined_bond_order}(a-d). The black regions above (green arrow in Fig.~\ref{fig7_confined_bond_order}(b)) and below (orange arrow) the sample (blue arrow) correspond to the top and bottom plates, respectively. We use confocal images to locate particle positions, and perform the bond order parameter analysis. 3D reconstructions of these particles can be seen in Fig.~\ref{fig7_confined_bond_order}(e-h). The particles with high $N_{\rm{ord}}$ are in blue, and low $N_{\rm{ord}}$ are in red. By inspection we observe that the local structural order varies non-monotonically with gap with crystals at intermediate gaps $h\approx 10(2a)$ appearing more ordered. In addition, the crystal structure appears to break up into smaller domains at small gaps.    

We quantify these observed trends by plotting the $N_{\rm{ord}}$ histograms for the four different gap heights $h$ in Fig.~\ref{fig7_confined_bond_order}(i-l). We find that for large gaps $h>10(2a)$ the distributions look similar with a relatively narrow width. As the sample is confined to smaller gaps, however, we observe that the distribution becomes broader with a larger probability of having particles with lower order. 
These behaviors are further quantified in Fig.~\ref{fig8_ordered_number}, in which the mean and standard deviation of $N_{\rm{ord}}$ for all seven gaps investigated are plotted versus $h/(2a)$. These measurements indicate that reduction in structural order and increase in its variance become more pronounced for gaps smaller than ten particle diameters. These observations are consistent with a large body of previous research indicating confinement effects become significant in samples confined to gaps containing less than ten layers~\cite{pieranski1983thin, cohen2004shear, neser1997finite, fortini2006phase}.

\subsection{Stress}

\begin{figure} [htp]
\includegraphics[width=0.45 \textwidth]{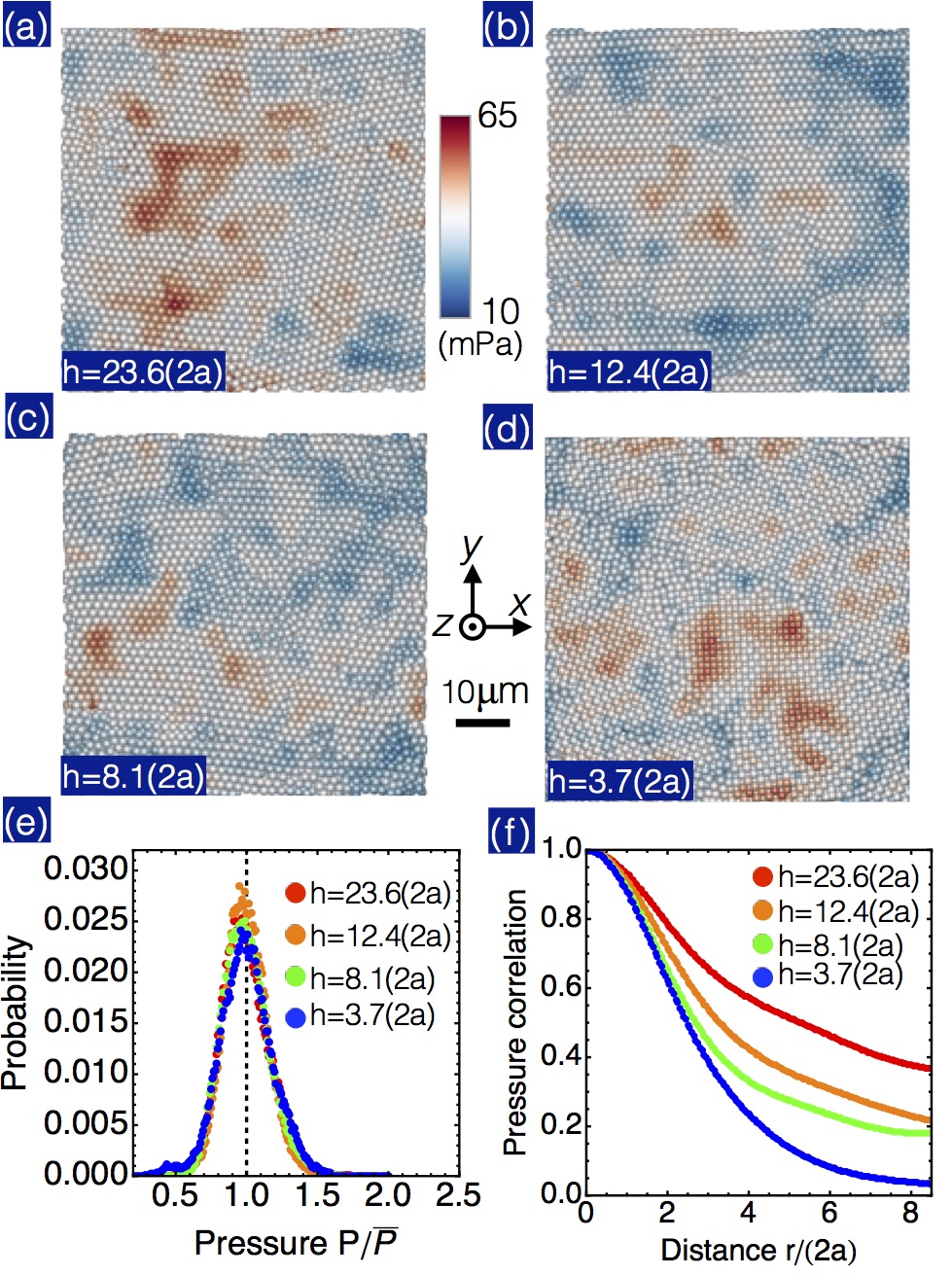}
\caption{Pressure distributions in confined polycrystals. (a-d) Horizontal slices ($\hat x \hat y$) of 3D pressure distributions in polycrystals for four representative $h$. The color scale is located between (a) and (b) and shared by all plots. (e) Histograms of normalized pressure $P/\bar P$ where $\bar P$ is the mean pressure, remain constant for all $h$. (f) Radial part of the pressure correlation function $C_P(r/2a)$. The correlation function decays faster when the system is confined indicating a shorter correlation length and pressure fluctuation localization.}
\label{fig9_pressure_distribution_confinement}
\end{figure}

We calculate the 3D pressure distributions of all tested samples, and show horizontal 2D slices ($\hat x \hat y$ at $z \approx h/2$) for the same four representative gap heights $h$ in Fig.~\ref{fig9_pressure_distribution_confinement}(a-d). Similar to the bulk measurements, the pressure fluctuations in confined samples are evenly spread, and roughly uncorrelated with the grain boundary arrangement. Furthermore, as anticipated, the mean pressure remains approximately constant (to within $10\%$ of the mean) as $h$ decreases. The constant pressure results from the fact that the confined sample is surrounded by a suspension reservoir residing beyond the confining plates that regulates the overall pressure of the confined zone. We also find that the standard deviation of the pressure is independent of $h$. The uniformity of the standard deviation is illustrated by the histograms of normalized pressure $P/\bar P$ shown in Fig.~\ref{fig9_pressure_distribution_confinement}(e), in which all four datasets collapse on a single Gaussian distribution, consistent with the Gaussian pressure distribution found in the bulk crystal samples (see Fig.~\ref{afig2_bulk_pressure}(d)). We note that while the structural order is substantially smaller at small gaps, the corresponding pressure histogram remains Gaussian. In particular, we do not see evidence of significant deviations from Gaussian behavior as is typically observed in glassy systems~\cite{bi2015statistical, corwin2005structural, tighe2008entropy, tighe2010force}. 

Although the mean and standard deviations of the pressure are unaffected by confinement, its spatial distribution substantially changes. In the bulk sample, we find that fluctuations are long-ranged $\sim$10 particles. However, under confinement, these fluctuations become localized, see (Fig.~\ref{fig9_pressure_distribution_confinement}(d)). Such a stress localization is characterized by calculating the correlation function $C_P(\vec r/2a) = \langle (P(\vec x+\frac{\vec r}{2a}) - \bar P) (P(\vec x) - \bar P) \rangle _{\vec x}/P_{var}$. Here, $C_P(\vec r/2a)$ is unity at origin $\vec r/2a = 0$ and zero at $\vec r/2a \to \infty$ while $P_{var}$ is the variance of the pressure. For simplicity, we plot the radial part of the correlation function $C_P(r/2a)$ in Fig.~\ref{fig9_pressure_distribution_confinement}(f). As shown, the correlation function decays faster for smaller gap size $h$, indicating a more localized pressure fluctuation. At the smallest gap we explored $(h/2a=3.7)$ the correlation length is $\approx 3.1(2a)$ about three times shorter than in bulk sample. This reduction in correlation length tracks the decrease in grain size (also approximately a factor of three) as the crystal is squeezed. 

\begin{figure} [htp]
\includegraphics[width=0.45 \textwidth]{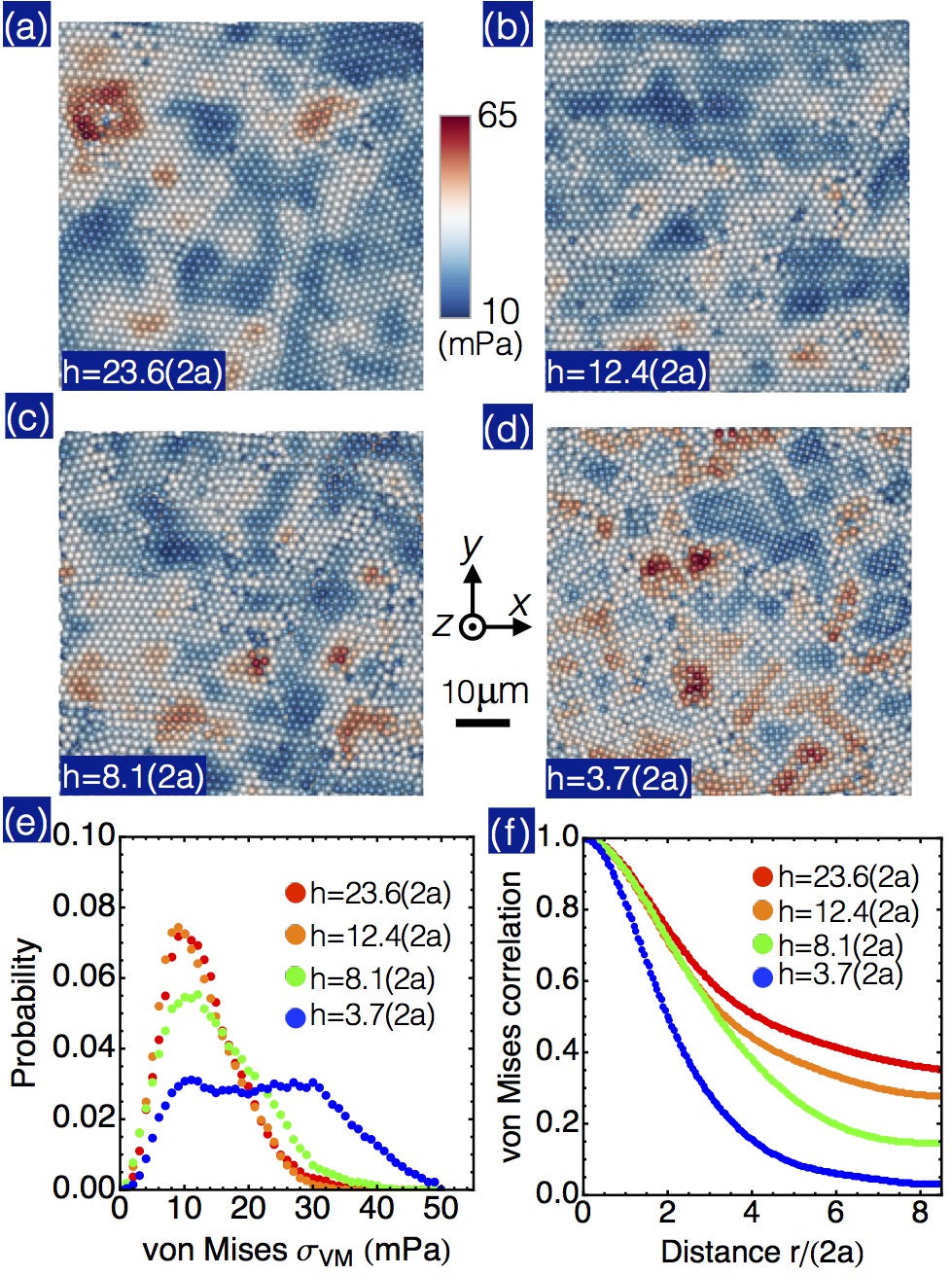}
\caption{von Mises stress $\sigma_{{\rm{VM}}}$ distributions in confined polycrystals. (a-d) Horizontal slices ($\hat x \hat y$) of 3D distributions for four representative $h$. The color scale is located between (a) and (b) and shared by all plots. (e) Histograms of $\sigma_{{\rm{VM}}}$. We find that the histogram extends toward the right indicating an enhancement of mean $\sigma_{{\rm{VM}}}$ when $h$ decreases. (f) Radial part of the von Mises stress correlation function $C_{\rm{VM}}(r/2a)$. The correlation function decays faster when the system is confined, indicating a  localized $\sigma_{{\rm{VM}}}$ fluctuation.}
\label{fig10_vonMises_distribution_confinement}
\end{figure}

We perform the same analysis on deviatoric stress for different gap sizes, and show 2D slices of the von Mises stress $\sigma_{{\rm{VM}}}$ fields for four different $h$, see Fig.~\ref{fig10_vonMises_distribution_confinement} (a-d). Consistent with the pressure distribution, the correlation between $\sigma_{{\rm{VM}}}$ and grain boundary arrangement is negligible. As shown in Fig.~\ref{fig10_vonMises_distribution_confinement}(e), the histogram of $\sigma_{{\rm{VM}}}$ shifts to the right and broadens with decreasing $h$, indicating a higher mean value and standard deviation. These trends become more pronounced when $h\le 10(2a)$. 

Similar to our treatment of the pressure, we quantify the length scale of $\sigma_{{\rm{VM}}}$ fluctuations by calculating its correlation function, $C_{\rm{VM}}(\vec r/2a) = \langle (\sigma_{{\rm{VM}}}(\vec x+\frac{\vec r}{2a}) - \bar \sigma_{{\rm{VM}}}) (\sigma_{{\rm{VM}}}(\vec x) - \bar \sigma_{{\rm{VM}}}) \rangle _{\vec x}/\sigma_{{\rm{VM}} var}$. We plot the radial component of the correlation $C_{\rm{VM}}(r/2a)$ in Fig.~\ref{fig10_vonMises_distribution_confinement}(f). Consistent with the pressure correlation evolution, $C_{\rm{VM}}(r/2a)$ also decays more rapidly with decreasing $h$. As with the pressure, this reduction in correlation length also tracks the decrease in grain size as the crystal is squeezed.

\begin{figure} [htp]
\includegraphics[width=0.3 \textwidth]{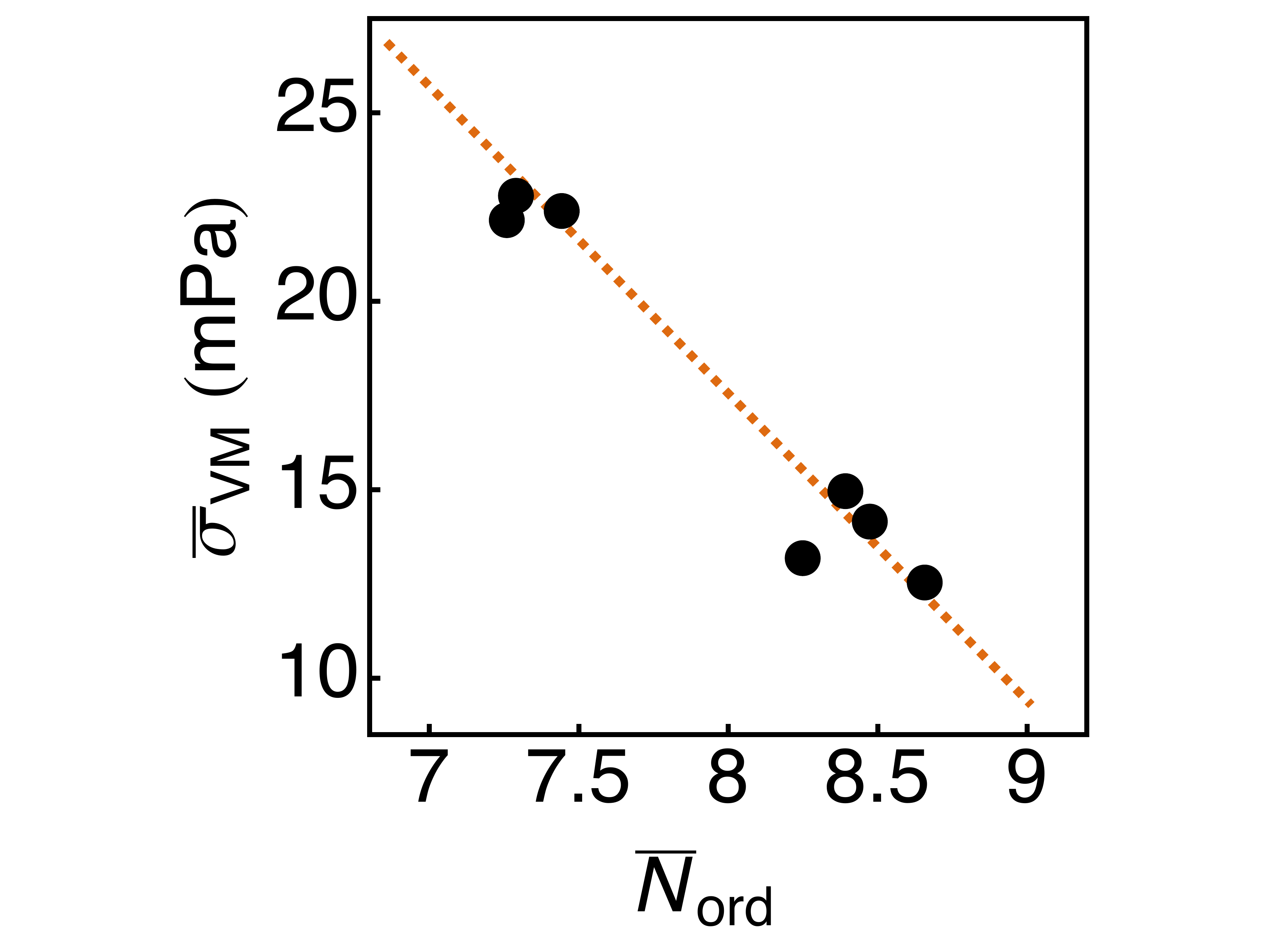}
\caption{Mean value of von Mises stress $\bar \sigma_{\rm{VM}}$ as a function $\bar N_{\rm{ord}}$. $\bar \sigma_{\rm{VM}}$ nearly doubles within a 15\% decrease in $\bar N_{\rm{ord}}$. The orange dashed line fits to the data denoting a substantial correlation between the structure and resulting deviatoric stress.}
\label{fig11_vonMises_order_confinement}
\end{figure}

The pronounced change in the $\sigma_{{\rm{VM}}}$ suggests a link between the local structure order and $\sigma_{{\rm{VM}}}$. To illustrate this relationship, we plot the mean deviatoric stress $\bar \sigma_{{\rm{VM}}}$ against the ordered neighbor number averaged over sample $\bar N_{\rm{ord}}$ in Fig.~\ref{fig11_vonMises_order_confinement}. The straight dashed line fits to the data showing a clear correlation between $\bar \sigma_{{\rm{VM}}}$ and $\bar N_{\rm{ord}}$. More importantly, the mean deviatoric stress approximately doubles while $\bar N_{\rm{ord}}$ decreases only by $\sim15\%$. In atomic systems higher $\bar \sigma_{{\rm{VM}}}$ indicates the system is closer to yielding suggesting a lower shear stress would be required to induce plastic deformation of the crystal.

\section{Discussion and Conclusions}
Our studies, which combine both structure and stress measurement at the single particle scale, clearly illustrate that confinement can have profound consequences for stress distributions in hard sphere crystals. While we find that the distribution of the pressure and the von Mises stress are weakly correlated with the grain boundary location, their fluctuations become more localized due to the additional grain boundaries introduced by our confining procedure. Thus, the grain boundaries effectively shield each grain from the specific spatial distribution of stresses in neighboring grains. 

In addition, we find that the deviatoric stress has a much wider distribution of magnitudes and a mean value that nearly doubles as the gap is reduced from $h/2a \approx 24$ to $h/2a \approx 4$. In our system, this reduction in gap produces smaller grain sizes. Such a grain size reduction has been suggested to have a great influence on the microstructure\cite{de2007grain}, defect density\cite{rose1997instability, wu2005mechanical}, and particle diffusivity~\cite{harrison1961influence} in polycrystals. More importantly, in atomic nanocrystals, grain size reduction leads to substantially lower yield stress a phenomena known as the inverse Hall-Petch relation~\cite{schiotz2003maximum, schiotz2004atomic}. Such trends are consistent with our observed enhancement in the deviatoric stress -- typically a measure of how close a system is to yielding. Further experiments in which the normal force is continuously measured or experiments in which the yield stress under shear is determined for crystals under different degrees of confinement would shed light on whether a direct link can be made between our colloidal system and films comprised of nanocrystalline grains.

Such studies would also help determine whether different yielding mechanisms dominate when the grain size is reduced. For example, it has been shown in numerical simulations that in large grains dislocations penetrate the grains and entangle resulting in strain hardening. In contrast, for small grains, plastic flow induces stacking faults and twining that localize near grain boundaries, and do not contribute significantly to the flow stress. Experiments in which we can simultaneously plastically deform the crystals while measuring their order and stress evolutions would elucidate whether similar mechanisms are at play in these hard sphere colloidal crystals.  

More broadly, combining the bulk stress measurement and SALSA provides a direct way to quantify the interplay between microscopic defect structures and macroscopic mechanical properties. This approach opens the door to uncovering the mechanisms that underly many defect-dominated phenomena in solid mechanics including defect-assisted premelting, strain hardening, and material fatigue.

\section{Appendix A: Crystal sample preparation and imaging details}

To grow a bulk crystal, we set the solvent density value ($\sim 1.20$ g/cm$^3$) slightly lower than the particle density ($\sim 1.23$ g/cm$^3$), so the particles can sediment and form a crystal with a higher volume fraction. We load a suspension with a volume fraction $\phi\approx 0.60$ in a sample cell and hermetically seal it. Prior to the experiment, the sample is placed on the microscope stage for at least 24 hours, until the sedimentation is complete. The resulting crystal has a total thickness of approximately 280 $\mu$m and volume fraction $\phi\approx$ 0.67 $\pm$ 0.03. During the crystal growth, the P$\acute{\rm{e}}$clect number for sedimentation is $\rm{Pe}_g = \Delta \rho g a^4/k_B T \sim 0.03$, where $\Delta \rho = 0.03$ g/cm$^3$ refers to the particle-solvent density mismatch, $g$ the gravitational acceleration, and $k_B T$ the thermal energy. The small P$\acute{\rm{e}}$clect number, $\rm{Pe}_g \ll 1$, indicates that the particle sedimentation rate is much slower than its self-diffusion. As a result, the PMMA sample can form polycrystals with far larger grain size and lower defect density than silica systems~\cite{lin2016measuring}, where the particle density is significantly mismatched. 

Our confined crystal samples are all density-matched, and we did not observe any significant particle sedimentation during the course of the experiment. The confined suspensions have a volume fraction $\phi \approx 0.63$, slightly lower than the bulk crystal value $\phi\approx$ 0.67, and have a lower viscosity, allowing it to be loaded in a confocal rheoscope~\cite{cohen2004shear}.

At an imaging rate 60~frames per second, the acquisition time of one confocal image stack that consists of 400 slices is $\sim$ 6.7~s. This time scale is comparable to a particle's relaxation time $\tau_D = 6 \pi \eta_0 a^3/k_B T\sim 4.8$~s where $\eta_0\sim 2.1$~mPa.s is the solvent viscosity. The final SALSA stress field is averaged over 15 image stacks requiring $\sim 100$~s to collect. We observe short-time stress fluctuations arising from particle random motions within their local environment over the image acquisition time. Our data is focused on these relatively short time scales rather than the stresses associated with the long time annealing of crystal grains~\cite{korda2001annealing} or the glassy behavior near grain boundaries~\cite{nagamanasa2011confined}.

\section{Appendix B: Bulk crystal structure}

\begin{figure} [htp]
\includegraphics[width=0.45 \textwidth]{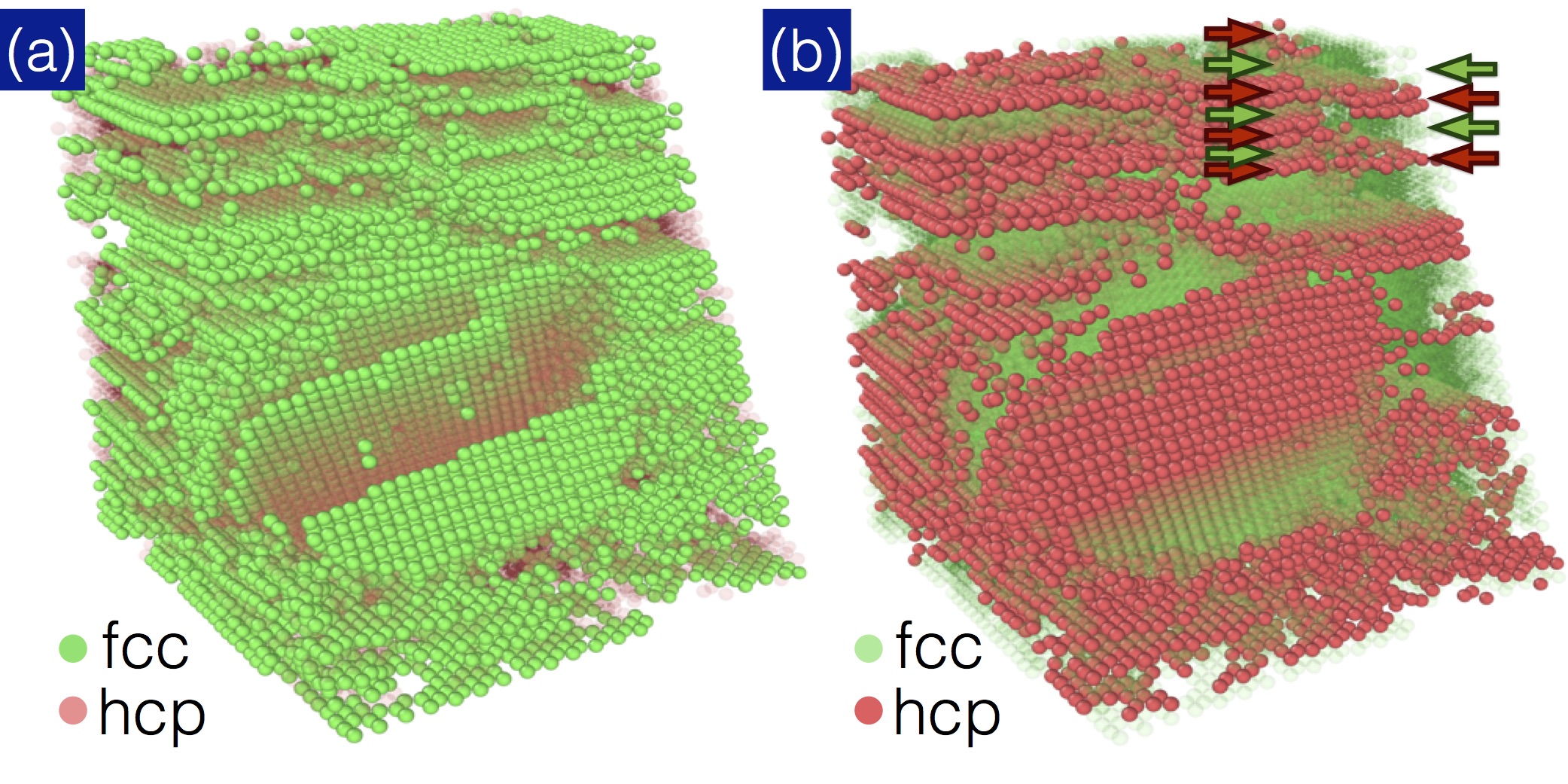}
\caption{We use OVITO~\cite{stukowski2009visualization} to visualize the fcc (green) and hcp (red) domains in (a) and (b). In (a) the fcc crystals are brighter and hcp crystals are dimmer. The color tone is reversed in (b). Stacking faults can be seen at the interfaces between fcc and hcp within horizontal planes.}
\label{afig_1_stacking_faults}
\end{figure}

To visualize the orientations and types of crystals, we perform a bond-angle analysis distinguishing the fcc and hcp crystals~\cite{ackland2006applications, stukowski2009visualization}. We highlight the color of fcc crystals (green) and tone down the color of hcp crystals (red) in Fig.~\ref{afig_1_stacking_faults}(a). In Fig.~\ref{afig_1_stacking_faults}(b) we reverse the color contrast. We find the fractions of fcc ($\approx 60\%$) and hcp ($\approx 40\%$) crystals comparable. This finding is consistent with previous computational studies, in which the fcc configuration is shown to be entropically favored over hcp, but only by $10^{-3}k_BT$ per particle~\cite{solomon2006stacking, mau1999stacking, frenkel1984new}. Furthermore, the observed fcc fractions $\alpha\sim 0.6$ is also consistent with the values found in previous scattering~\cite{van1993expansion, pusey1989structure} and direct imaging~\cite{verhaegh1995direct, elliot1997direct} experiments, which show that colloidal polycrystals are essentially comprised of randomly-stacked hexagonal layers with $\alpha \approx 0.5$. We also find that the horizontal fcc and hcp stacks (with a crystal orientation 111 parallel to $z$-axis) do not alternate consistently across the field of view, as indicated by the arrows in Fig.~\ref{afig_1_stacking_faults}(b). This stacking inconsistency leads to stacking faults that can be identified by the vertical interfaces between the two crystals. The high density of these planar defects in our nearly-equilibrium crystals arises from the small free energy cost associated with the fault formation. By observing the crystal orientation, we also find that most of the crystal domains have a (111) plane parallel to the coverslip. This parallel alignment is consistent with the previously reported mechanism of crystal growth from sedimentation~\cite{ramsteiner2009experimental}. In this case, the first few crystal layers form simultaneously near the flat bottom plate, while further crystals grow layer by layer.

\section{Appendix C: Bulk crystal pressure}\label{bulk_pressure}

\begin{figure}[htp]
\includegraphics[width=0.4 \textwidth]{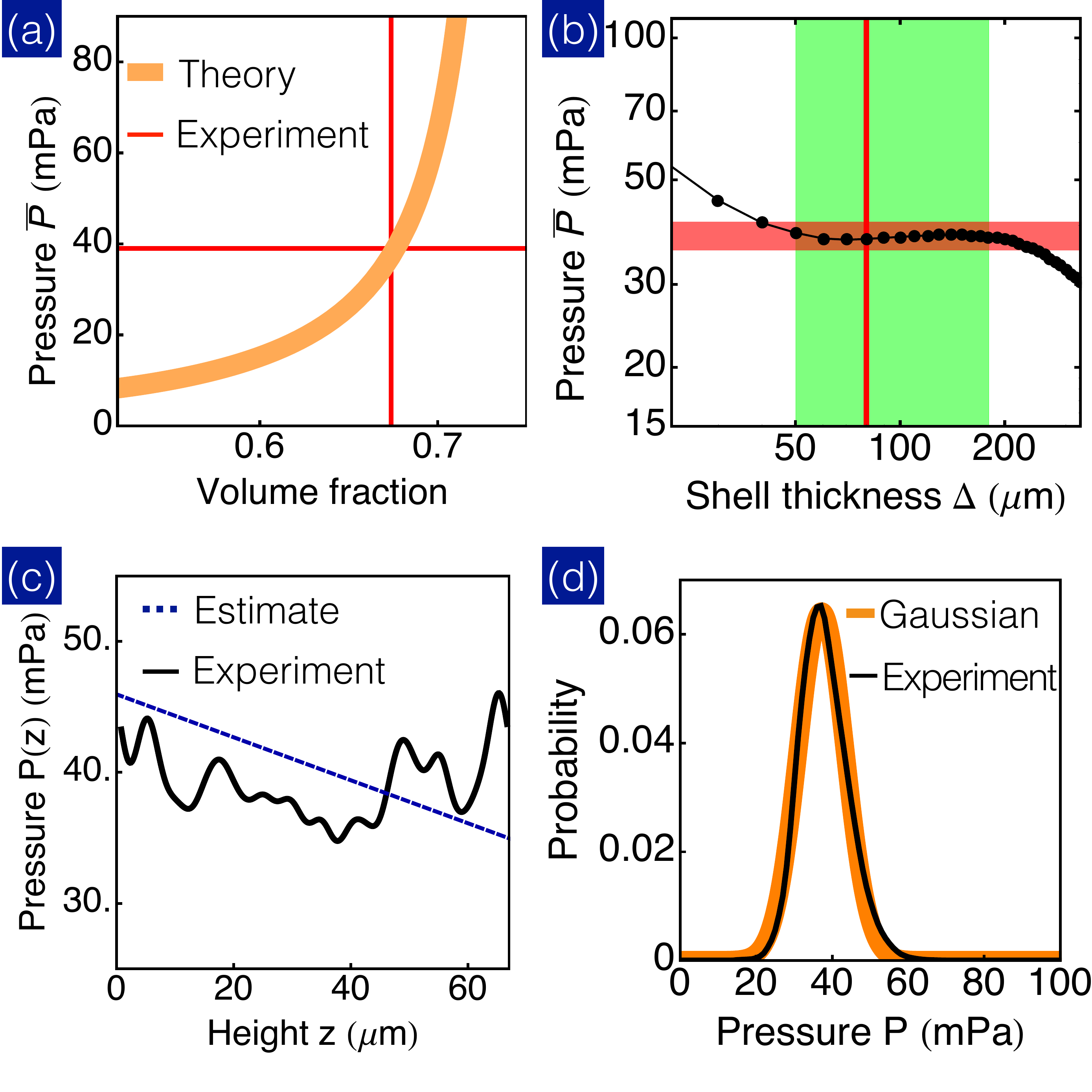}
\caption{Bulk pressure measured with SALSA. (a) The observed mean pressure 39 mPa (horizontal red line) is consistent with the hard-sphere simulation (orange line) at the volume fraction found in our system $\phi \sim 0.67$ (vertical red line). (b) The mean pressure is insensitive to the shell thickness $\Delta$ in the range 50 nm$\le \Delta \le$180 nm (green shade). The shell thickness used in the current work is $\Delta = 80$ nm (red vertical line). Horizontal line is a guide to the eye. (c) Pressure averaged in $\hat x \hat y$ plane is plotted as a function of height $z$. The blue dashed line is the estimated hydrostatics pressure. (d) The pressure histogram (black) is consistent with Gaussian distribution (orange) with a standard deviation about 15\% to the mean.}
\label{afig2_bulk_pressure}
\end{figure}

We find excellent agreement between our measured pressure value (horizontal line in Fig.\ref{afig2_bulk_pressure}(a)) and the value predicted by simulations (orange line) of hard spheres at the volume fraction found ($\phi \approx$ 0.67, vertical line) in our system. To investigate how this pressure value depends on the SALSA shell thickness $\Delta$ , we plot $\bar P$ versus $\Delta$ (black joined points) in Fig.~\ref{afig2_bulk_pressure} (b). The constant $\bar P$ highlighted by the green shade and horizontal red line indicates a pressure value insensitive to shell thickness between 50 nm $\le \Delta \le$ 180 nm. The overestimated pressure at $\Delta<$ 50 nm arises from polydispersity and the particle overlap associated with featuring uncertainties. The underestimated pressure at $\Delta>$ 180 nm arises from the saturation in the collision probability once all the nearest neighbors are included. The observed $\Delta-$independent pressure confirms that both our imaging and particle featuring resolutions are adequate to quantify the particle collision probability and its resulting stress. Throughout all SALSA analyses in this work, we set the shell thickness $\Delta = 80$ nm.

The particle polydispersity has different influences on the normal and shear stress measurements. A detailed discussion of this issue can be found in our previous work (Supplementary Information of~\cite{lin2016measuring}), where the shear and normal stresses near a vacancy defect are measured. Overall, since the pressure measurement relies on a more accurate identification of colliding particles, it can be influenced by the polydispersity more notably. In contrast to the pressure, the shear component is primarily related to the angular anisotropy of the neighboring particle configuration; hence, it is affected by the polydispersity less. In the current work, the particle polydispersity ($\le$3\%$\sim40$~nm comparable to the particle featuring error) is readily smaller than the shell thickness $\Delta=80$ nm that defines the colliding criterion. Therefore, we anticipate that the identification of colliding particle is primarily associated with $\Delta$ rather than polydispersity.

Finally, since we grow our bulk polycrystal sample from particle sedimentation, we anticipate the crystal pressure depends on the sample thickness and the mismatched density between particle and solvent. Furthermore, we anticipate that the pressure should not significantly vary with the height $z$, given that the depth of view $\approx$ 68 $\mu$m is much thinner than the sample thickness $z_c\approx$ 280 $\mu$m. We plot the pressure averaged over the $x-y$ plane as a function of height $z$ in Fig.~\ref{afig2_bulk_pressure}(c). Overall, the pressure trend is consistent with the estimated hydrostatic pressure arising from gravity, $\Delta \rho g (z_c - z) \phi$ (blue dashed line), and we do not observe a clear decay in pressure. The pressure variation is mainly dominated by the fluctuations from the defect distribution in the sample. In Fig.~\ref{afig2_bulk_pressure}(d) we show the histogram of the pressure fluctuation. The histogram can be described well by a Gaussian distribution (orange line) with a standard deviation $\approx 6$ mPa corresponding to 15$\%$ of the mean. This magnitude of pressure fluctuation is consistent with previous results found in silica systems~\cite{lin2016measuring}.

\begin{figure}[htp]
\includegraphics[width=0.48 \textwidth]{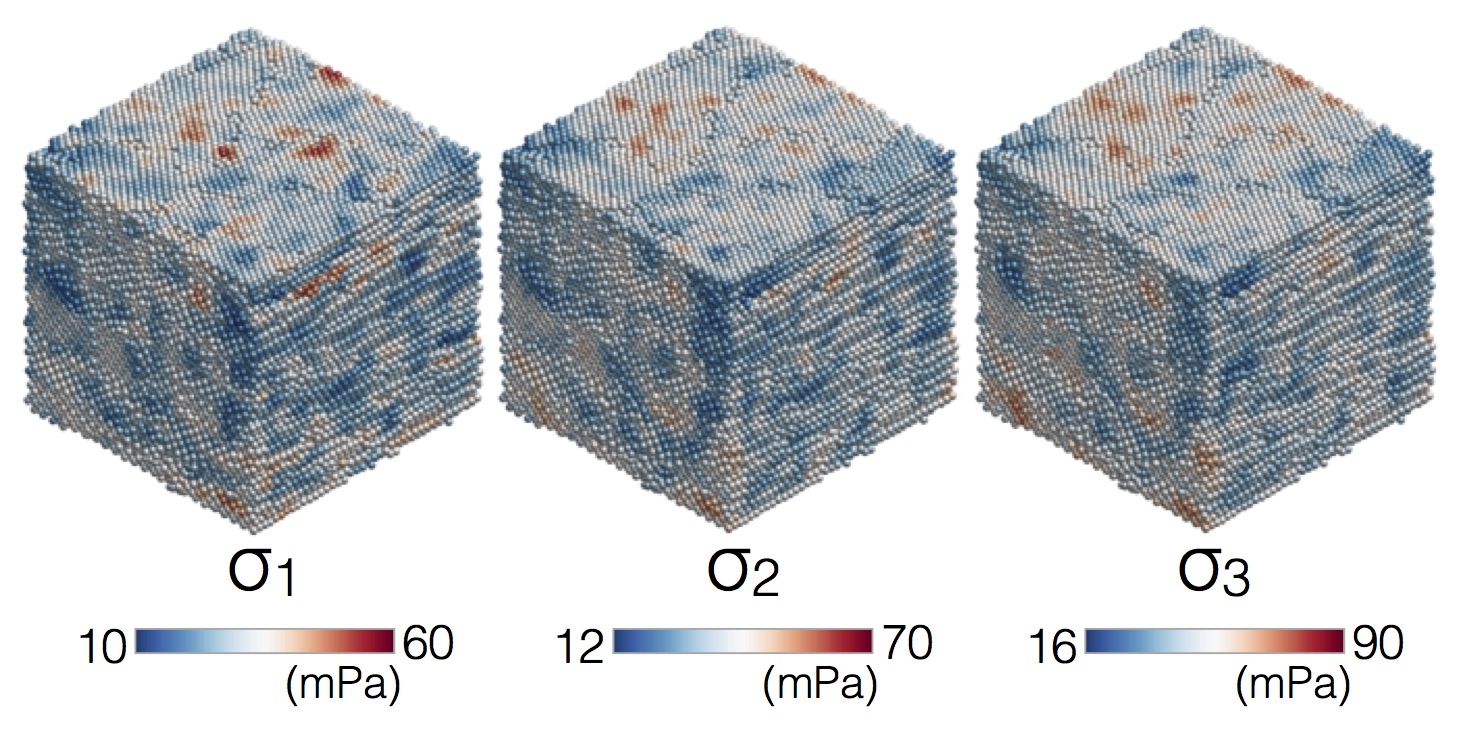}
\caption{3D principal stresses of the tested PMMA polycrystal. As illustrated by the color bars, blue particles denote low stresses and red particles denote high stresses.}
\label{afig_3_principal_stress}
\end{figure}

\begin{figure}[htp]
\includegraphics[width=0.37 \textwidth]{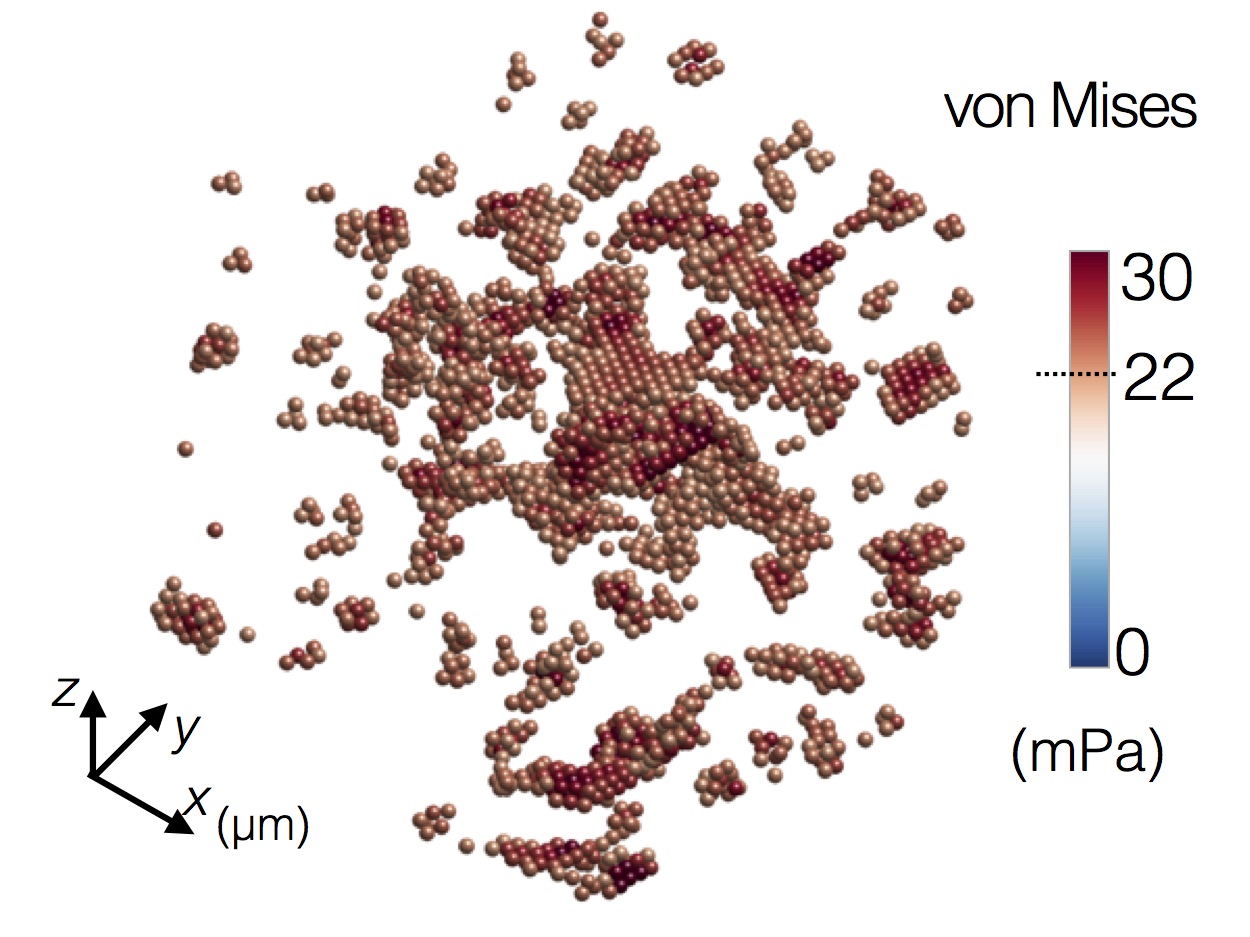}
\caption{The thresholded von Mises stress field $\sigma_{\rm{VM}}\ge$ 22 mPa. We find no significant correlation between the grain boundary particles and the particles with higher $\sigma_{\rm{VM}}$. The thresholded value is denoted by the horizontal dashed line on the color bar.}
\label{afig_4_threshold_vonMises}
\end{figure}

\section{Appendix D: Principal and von Mises stresses in bulk crystals}\label{bulk_principal_vonMises}
We calculate the three principal stresses $\sigma_1$, $\sigma_2$, and $\sigma_3$ and show them in Fig.~\ref{afig_3_principal_stress}. The three principal stresses are the eigenvalues of the stress tensor, and are thus independent of the coordinate orientation. The common trend between the three principal stresses corresponds to the hydrostatic pressure variation, while the difference is related to the deviatoric stress $\sigma_{\rm{VM}}$.

We calculate $\sigma_{\rm{VM}}$ and visualize its distribution within the sample by excluding particles with $\sigma_{\rm{VM}}$ < 22 mPa. The thresholded field is shown in Fig.~\ref{afig_4_threshold_vonMises}. We find no obvious correlation between the distribution of the particles with large $\sigma_{\rm{VM}}$ and the grain boundaries within the polycrystal. 

\section{Acknowledgements}
The authors thank Marc Miskin, Lena Bartell, Brian Leahy, Meera Ramaswamy, Eric Schwen, and Matthew Bierbaum for helpful conversations. This work was supported by NSF DMR-CMP Award No. 1507607.





\providecommand*{\mcitethebibliography}{\thebibliography}
\csname @ifundefined\endcsname{endmcitethebibliography}
{\let\endmcitethebibliography\endthebibliography}{}

\end{document}